\documentclass[onecolumn,10.5pt]{IEEEtran}
\usepackage{amsfonts,amsmath,amssymb,amsthm,caption,algorithmic}
\usepackage{amsbsy}
\usepackage{graphicx,multirow,bm}
\usepackage{color}
\makeatletter
\newtheoremstyle{mythm}{3pt}{3pt}{}{16pt}{\bfseries}{:}{.5em}{}
\theoremstyle{mythm}
\newtheorem{theorem}{Theorem}
\newtheorem{definition}{Definition}

\newtheorem{proposition}{Proposition}
\newtheorem{conjecture}{Conjecture}
\newtheorem{coro}{Corollary}
\newtheorem{lemma}{Lemma}

\begin{document}
\title{
Probabilistic Existence Results for Parent-Identifying Schemes
\\[0.5cm]
\author{Yujie Gu, Minquan Cheng, Grigory Kabatiansky, and Ying Miao
}
\thanks{Y. Gu is with the Department of Electrical Engineering--Systems, Tel Aviv University, Tel Aviv, Israel (e-mail: guyujie2016@gmail.com).
This work was done while the author was with Graduate School of Systems and Information Engineering, University of Tsukuba, Japan.}
\thanks{M. Cheng is with the Guangxi Key Lab of Multi-Source Information Mining and Security, Guangxi Normal University, Guilin 541004, China (e-mail: chengqinshi@hotmail.com). Research in part supported by 2016GXNSFFA380011, Guangxi Higher Institutions Program of Introducing 100 High-Level Overseas Talents, Research Fund of Guangxi Key Lab of Multi-source Information Mining $\&$ Security (16-B-01).}
\thanks{
G. Kabatiansky is with Skolkovo Institute of Science and Technology (Skoltech), Moscow, Russia (e-mail: g.kabatyansky@skoltech.ru).
}
\thanks{Y. Miao is with the Faculty of Engineering, Information and Systems, University of Tsukuba, Tsukuba, Ibaraki 305-8573, Japan (e-mail: miao@sk.tsukuba.ac.jp). Research supported by JSPS Grant-in-Aid for Scientific Research (B) under Grant No. 18H01133.}
}
\date{}
\maketitle

\vspace{0.1in}
\begin{abstract}
Parent-identifying schemes provide a way to identify causes from effects for some information systems such as digital fingerprinting and group testing.
In this paper, we consider combinatorial structures for parent-identifying schemes.
First, we establish an equivalent relationship between parent-identifying schemes and forbidden configurations.
Based on this relationship,
we derive probabilistic existence lower bounds for two related combinatorial structures, that is, $t$-parent-identifying set systems ($t$-IPPS) and $t$-multimedia parent-identifying codes ($t$-MIPPC), which are used in broadcast encryption and multimedia fingerprinting respectively.
The probabilistic lower bound for the maximum size of a $t$-IPPS has the asymptotically optimal order of magnitude in many cases,
and that for $t$-MIPPC
provides the asymptotically optimal code rate when $t=2$ and the best known asymptotic code rate when $t\ge 3$. Furthermore, we analyze the structure of $2$-IPPS and prove some bounds for certain cases.
\end{abstract}


\section{Introduction} %
\label{pre}                     %

The elucidation of cause-effect relationships among variables or events is the central aim of many studies, see preface of Pearl's book \cite{Pearl2009}. Parent-identifying schemes provide a way to identify causes from an effect for some information systems such as digital fingerprinting and group testing.

We use the terminology of \textit{$f$-channel} to represent a kind of cause-effect relationship, which has at most $t$ inputs (causes) and a collection of possible outputs (effects). Let $Q$ be a finite alphabet of cardinality $|Q|=q$, in which elements represent  possible causes. Let $Q^n=\{\mathbf{c}=(\mathbf{c}(1),\ldots,\mathbf{c}(n)):\, \mathbf{c}(i)\in Q\}$, and $2^Q$ denote the power set of $Q$. We define \textit{$f$-channel} as a function $f$ from subsets of some ground set, which will be $Q^n$ or $2^Q$ in our settings, to subsets of the same ground set. Two most interesting particular choices of $f$ will be described below.

\begin{definition}\label{defIPPscheme}
A subset $\mathcal{C}$ of $Q^n$ or $2^Q$ is a \textit{$t$-parent-identifying scheme} under $f$-channel if for any $\mathcal{C}'\subseteq \mathcal{C}$ with $|\mathcal{C}'|\le t$ and any $d\in f(\mathcal{C}')$, we have
\begin{equation}
\label{def}
\bigcap_{\mathcal{P}\subseteq \mathcal{C}:\ |\mathcal{P}|\le t,\, d\in f(\mathcal{P})} \mathcal{P}\ne \emptyset.
\end{equation}
\end{definition}

The above definition exhibits the essential idea of cause clarification algorithms in parent-identifying schemes, that is, once an effect $d$ is observed then at least one cause can be found which is common to all subsets $\mathcal{P}\subseteq \mathcal{C}$ with size at most $t$ capable of generating $d$. If $d$ can be generated by $\mathcal{P}$, then we call $d$ a \textit{descendant} of $\mathcal{P}$ and call $\mathcal{P}$ a \textit{possible parent set} of $d$.

A set $\mathcal{C}\subseteq Q^n$ is called a $q$-ary  \textit{code} of length $n$ and each $\mathbf{c}\in \mathcal{C}$ is called a \textit{codeword}. In Definition \ref{defIPPscheme}, if $\mathcal{C}\subseteq Q^n$, then it can be regarded as a class of fingerprinting codes used in digital copyright protection. Well-known examples include \textit{codes with the $t$-identifiable parent property ($t$-IPP codes)}
\cite{Alon2001}--\cite{Blackburn2003IPP, Hollmann1998,  Staddon2001, To2004}, where
\begin{equation}\label{eq-f-IPP}
f(\mathcal{P})=\text{desc}(\mathcal{P})=\mathcal{P}(1)\times\mathcal{P}(2)\times\cdots\times\mathcal{P}(n),
\end{equation}
and $\mathcal{P}(i)=\{\mathbf{c}(i):\ \mathbf{c}\in \mathcal{P}\}$ is the $i$th projection of the set $\mathcal{P}$.

In the case when $\mathcal{C}$ is a collection of subsets of $Q$, i.e., $\mathcal{C}\subseteq 2^{Q}$, we call $\mathcal{C}$ a set system.
The following particular example, namely, \textit{$t$-single-user tracing superimposed family}, was introduced for applications in molecular biology \cite{Alon2006, Csuros2005}, where
\begin{equation}
\label{OR}
f(\mathcal{P})=\left\{\bigcup_{A\in \mathcal{P}}A\right\}.
\end{equation}
The $f$-channel described by \eqref{OR} is known as OR-(multiple-access) channel and codes for this channel are called superimposed codes, which were introduced half a century ago in \cite{Sing}, see also \cite{Erd1}, \cite{Erd2}. Note that the aim of superimposed codes is to identify the whole set $\mathcal{P}$ by its output $\{d\}=f(\mathcal{P})$, and the aim of $t$-single-user tracing superimposed family is to identify at least one element which belongs to all $\mathcal{P}$ such that $\{d\}=f(\mathcal{P})$, where $d=\bigcup_{A\in \mathcal{P}}A$. For {\it $t$-parent-identifying set systems ($t$-IPPS)} considered in this paper, the corresponding $f$-channel is described by
\begin{equation}
\label{IPPS}
f(\mathcal{P})=\left\{B: |B|\geq w, B\subseteq \bigcup_{ A\in \mathcal{P}}A\right\}.
\end{equation}
In some sense, this $f$-channel is  more complicated than the channel \eqref{OR} because its output is not a single subset but a family of subsets.

In this paper, we investigate parent-identifying set systems (IPPS) for broadcast encryption and multimedia parent-identifying codes (MIPPC) for multimedia fingerprinting. We focus on their (maximum) sizes (or in other words, code rates), one of the most important parameters, of such parent-identifying schemes. The best known upper bounds for the size of $t$-IPPS and $t$-MIPPC were proved in \cite{GM2016} and \cite{Cheng2017} respectively. In the literature, there are no general lower bounds for the maximum size of $t$-IPPS and $t$-MIPPC. In this paper, we will provide probabilistic constructions for $t$-IPPS and $t$-MIPPC, and compare their cardinalities with the known upper bounds respectively.

To that end, first, we establish an equivalent relationship between parent-identifying schemes and forbidden configurations. Based on this relationship, we derive probabilistic existence results for IPPS and MIPPC respectively. Accordingly, the probabilistic lower bound for $t$-IPPS has the asymptotically optimal order of magnitude in many cases, and that for $t$-MIPPC provides the asymptotically optimal code rate when $t=2$ and the best known asymptotic code rate when $t\ge 3$. Furthermore, we analyze the structure of $2$-IPPS and prove some bounds for certain cases.

The paper is organized as follows. First, in Section \ref{forbidden}, we state the definition of forbidden configuration, and establish an equivalent relationship between it and parent-identifying schemes. In Section \ref{IPPS}, we provide the probabilistic construction for $t$-IPPS and derive some bounds for $2$-IPPS with small $w$. We show the probabilistic existence result for $t$-MIPPC in Section \ref{MIPPC}. Finally, we conclude this paper in Section \ref{conclusion}.

\section{Forbidden configurations}
\label{forbidden}                     %
\subsection{Configuration}
In \cite{Barg2001}, Barg \textit{et al.} exploited the notion of \textit{minimal forbidden configuration} to study codes with the $t$-identifiable parent property.
We recap the related notions here.

\begin{definition}\label{defconfiguration}
In a set $\mathcal{C}$, let $\mathcal{F}=\{\mathcal{F}_1,\ldots,\mathcal{F}_m\}$ be a collection of subsets of $\mathcal{C}$ with $\mathcal{F}_i\subseteq \mathcal{C},\ |\mathcal{F}_i|\le t$, $i=1,\ldots,m$. Then $\mathcal{F}$ is called a \textit{configuration} if it has an empty intersection, i.e. $\bigcap_{1\le i\le m}\mathcal{F}_i=\emptyset$. Moreover, $\mathcal{F}$ is called a \textit{minimal configuration} if it is minimal under inclusion, that is, \begin{equation*}
\bigcap_{1\le j\le m,\atop {j\ne i}}\mathcal{F}_j\ne \emptyset, \ \ \forall \, 1\le i\le m.
\end{equation*}
\end{definition}
Denote $U(\mathcal{F})=\bigcup_{1\le i\le m}\mathcal{F}_i$. The cardinality of $U(\mathcal{F})$ is called the \textit{size} of the configuration $\mathcal{F}$. The following lemma shows that the size of a minimal configuration cannot be too large, which was shown in \cite{Barg2001}, \cite{Staddon2001}. To be self-contained, we expose its proof here. Denote $u:=\lfloor(\frac{t}{2}+1)^2\rfloor$.

\begin{lemma}[\cite{Barg2001}, \cite{Staddon2001}]\label{sizeminimalconf}
Let $\mathcal{F}$ be a minimal configuration. Then $|U(\mathcal{F})|\le u$.
\end{lemma}

\begin{IEEEproof}
Suppose $\mathcal{F}=\{\mathcal{F}_1,\ldots,\mathcal{F}_m\}$ be a minimal configuration. Then for each $1\le i\le m$, there exists a codeword $\mathbf{x}_i$ such that
\begin{equation*}
\mathbf{x}_i\notin \mathcal{F}_i\ \ \text{and}\ \ \mathbf{x}_i\in \bigcap_{1\le j\le m,\atop j\ne i} \mathcal{F}_j.
\end{equation*}
Clearly, $\mathbf{x}_i\ne \mathbf{x}_j$ for all $1\le i<j\le m$. Therefore
\begin{equation*}
\begin{split}
|U(\mathcal{F})|&=|\{\mathbf{x}_1,\mathbf{x}_2,\ldots,\mathbf{x}_m\}|+\left|\bigcup_{1\le i\le m}\mathcal{F}_i\setminus \{\mathbf{x}_1,\ldots,\mathbf{x}_m\}\right|\\
&\le m+\sum_{1\le i\le m}(|\mathcal{F}_i|-(m-1))\\
&\le m+m(t-m+1)\\
&=-m^2+(t+2)m\\
&\le (\frac{t}{2}+1)^2.
\end{split}
\end{equation*}
where the last inequality holds by taking $m=\frac{t}{2}+1$. The lemma follows.
\end{IEEEproof}

\subsection{Relationship}
To establish the relationship between a parent-identifying scheme and a (minimal) configuration, first, we define a forbidden configuration as follows.

\begin{definition}\label{defforbiddenconfiguration}
Let $\mathcal{C}$ be a $t$-parent-identifying scheme under $f$-channel. A (minimal) \textit{forbidden} configuration in $\mathcal{C}$ is a (minimal) configuration $\mathcal{F}=\{\mathcal{F}_1,\ldots,\mathcal{F}_m\}$ such that
\begin{equation*}
f(\mathcal{F}_1)\cap f(\mathcal{F}_2)\cap\cdots\cap f(\mathcal{F}_m)\ne \emptyset.
\end{equation*}
\end{definition}

The following relationship is from Lemma 1 and Definitions 1 and 3.

\begin{proposition}\label{relationshipIPPscheme}
A set $\mathcal{C}$ is a $t$-parent-identifying scheme under $f$-channel if and only if there are no minimal forbidden configurations in $\mathcal{C}$ with size at most $u$.
\end{proposition}

\begin{IEEEproof}
Indeed, if $\mathcal{C}$ is  a $t$-parent-identifying scheme, then by the definition there are no forbidden configurations and hence no minimal forbidden configurations. If $\mathcal{C}$ is  not a $t$-parent-identifying scheme then  there exists a subset $\mathcal{C}'\subseteq \mathcal{C}$, $|\mathcal{C}'|\le t$, and a descendant $d\in f(\mathcal{C}')$ such that
\begin{equation*}
\bigcap_{\mathcal{P}\in P_t(d)} \mathcal{P}= \emptyset.
\end{equation*}
Then we can find a minimal forbidden configuration in
\begin{equation*}
P_t(d)=\{\mathcal{P}\subseteq \mathcal{C}:\ |\mathcal{P}|\le t,\, d\in f(\mathcal{P})\}.
\end{equation*}
This is doable since we can consecutively remove some $\mathcal{P}$ from $P_t(d)$, if the intersection of remaining subsets is still empty, until it forms a minimal configuration. Then according to Lemma \ref{sizeminimalconf}, the size of any minimal configuration is at most $u$. The proof is completed.
\end{IEEEproof}

By Proposition \ref{relationshipIPPscheme}, we immediately have

\begin{coro}
A set $\mathcal{C}$ is a $t$-parent-identifying scheme under $f$-channel if and only if every subset $\mathcal{C}'\subseteq \mathcal{C}$ such that $|\mathcal{C'}|\le u$ is a $t$-parent-identifying scheme under $f$-channel.
\end{coro}

Notice that different $f$-channels lead to different parent-identifying schemes. By Proposition \ref{relationshipIPPscheme}, different parent-identifying schemes correspond to different minimal forbidden configurations. To derive a lower bound for the maximum size of a parent-identifying scheme, we can use the method of random coding with expurgation, that is, to estimate the expectation number of its corresponding minimal forbidden configurations, and then delete one element from each of them to destroy these minimal forbidden configurations. In the following sections, we will exhibit probabilistic existence results for IPPS and MIPPC in this way.

\section{Bounds for IPPS}
\label{IPPS}
\subsection{Parent-identifying set system}
In broadcast encryption, the distributor broadcasts the encrypted data and sends a valid key to each authorized user. To protect the copyright of broadcasted contents, the distributor will send different users with distinct keys, which represent their own identification. In fact, the authorized keys are essentially the same with the embedded fingerprints in digital fingerprinting. Both of them aim to resist collusion attacks. This kind of key-distributing schemes has already been investigated, see \cite{Chor1994}--\cite{Collins2009,GM2016,Stinson1998} for example.

Suppose the distributor has a set of base keys $\mathcal{X}$ with size $v$. We also call an element $x\in \mathcal{X}$ a \textit{point}. According to \cite{Stinson1998} an authorized user can receive $w(\le v)$ distinct base keys from the distributor and use it to decrypt the broadcast-encrypted contents based on a threshold secret sharing scheme \cite{Bob}, \cite{Shamir}. Hence the set of all authorized users' fingerprints is a subset $\mathcal{B}\subseteq \binom{\mathcal{X}}{w}$, where $\binom{\mathcal{X}}{w}$ denotes the set of all $w$-subsets of $\mathcal{X}$ and each authorized user $b$ receives its own $w$-subset $B\in \mathcal{B}$. The pair $(\mathcal{X},\mathcal{B})$ is called a $(w,v)$ \textit{set system}, and each $B\in \mathcal{B}$ is called a \textit{block}. It is often assumed that at most $t$ dishonest authorized users $\mathcal{B}'\subseteq \mathcal{B}$ would collude to generate a pirate fingerprint $T$ such that $T\subseteq \bigcup_{B\in \mathcal{B}'}B$ and $|T|=w$, that is, the corresponding {\it set system's} $f$-channel is defined as
\begin{equation*}
f_{ss}(\mathcal{B}')=\left\{T\subseteq \mathcal{X}:\ |T|=w,\ T\subseteq \bigcup_{B\in \mathcal{B}'}B \right\}.
\end{equation*}
To resist this kind of attack, the parent-identifying set systems were defined in \cite{Collins2009}.

\begin{definition}\label{defIPPS}
A \textit{$t$-parent-identifying set system}, denoted as $t$-IPPS$(w,v)$, is a pair $(\mathcal{X},\mathcal{B})$ such that $|\mathcal{X}|=v$, $\mathcal{B}\subseteq \binom{\mathcal{X}}{w}$, with the property that for any $w$-subset $T\subseteq \mathcal{X}$, either $P_t(T)$ is empty, or
\begin{equation*}
\bigcap_{\mathcal{P}\in P_t(T)}\mathcal{P}\ne \emptyset,
\end{equation*}
where
\begin{equation*}
P_t(T)=\left\{\mathcal{P}\subseteq \mathcal{B}:\ |\mathcal{P}|\le t,\ T\subseteq \bigcup_{B\in \mathcal{P}}B\right\}.
\end{equation*}
\end{definition}

The number of blocks $B \in \mathcal{B}$ is called the \textit{size} of this $t$-IPPS$(w,v)$. Denote $I_t(w,v)$ as the maximum size of a $t$-IPPS$(w,v)$. A $t$-IPPS$(w,v)$ $(\mathcal{X},\mathcal{B})$ is called \textit{optimal} if it has size $I_t(w,v)$.

\subsection{A lower bound for $t$-IPPS}

According to Definition \ref{defforbiddenconfiguration}, we have the following description of the forbidden configuration in an IPPS. Let $(\mathcal{X},\mathcal{B})$ be a $t$-IPPS$(w,v)$. A \textit{(minimal) forbidden configuration} $\mathcal{F}=\{\mathcal{F}_1,\ldots,\mathcal{F}_m\}$ in $\mathcal{B}$ is a (minimal) configuration in $\mathcal{B}$ such that
\begin{equation*}
\left|\left(\bigcup_{B\in \mathcal{F}_1}B\right)\cap \left(\bigcup_{B\in \mathcal{F}_2}B\right)\cap \cdots \cap \left(\bigcup_{B\in \mathcal{F}_m}B\right)\right|\ge w.
\end{equation*}

Therefore we have the following lemma in this setting.
\begin{lemma}\label{pointsIPPS}
If a minimal forbidden configuration $\mathcal{F}=\{\mathcal{F}_1,\ldots,\mathcal{F}_m\}$  contains $s$ distinct blocks, where $2\le s \le u$, then it is spanned by at most $(s-1)w$ points, that is,
\begin{equation*}
\left|\bigcup_{B\in U(\mathcal{F})}B\right|=\left|\bigcup_{1\le i\le m}(\bigcup_{B\in \mathcal{F}_i}B)\right|\le { \min}\{(s-1)w,v \}.
\end{equation*}
\end{lemma}

\begin{IEEEproof}
Suppose $\mathcal{F}=\{\mathcal{F}_1,\ldots,\mathcal{F}_m\}$ is a minimal forbidden configuration with size $|U(\mathcal{F})|=s$. Then there exists a $w$-subset $W\subseteq \mathcal{X}$ such that
\begin{equation*}
W\subseteq \bigcup_{B\in \mathcal{F}_i}B,\ \ \forall\, 1\le i\le m.
\end{equation*}
Thus each point in $W$ appears in at least two distinct blocks of $U(\mathcal{F})$, since if not, it contradicts that $\mathcal{F}$ is a configuration. If one counts points with multiplicities in all blocks $B$ belonging to the configuration $\mathcal{F}$, then the corresponding sum equals to $sw$ points. Since all points of $W$ were counted at least twice in this sum, the number of distinct points in $\bigcup_{B\in U(\mathcal{F})}B$ is at most $sw-w$. The lemma follows.
\end{IEEEproof}

Combining Proposition \ref{relationshipIPPscheme} and Lemma \ref{pointsIPPS}, we have
\begin{coro}\label{sufficientIPPS}
If a set system $(\mathcal{X},\mathcal{B})$ is not a $t$-IPPS$(w,v)$, then there exists an $s$-subset $\mathcal{U}\subseteq \mathcal{B}$, $2\le s\le u$, such that $\mathcal{U}$ is spanned by at most $(s-1)w$ points.
\end{coro}

For convenience, we introduce the notion of a {\it bad $s$-packet} as follows.
\begin{definition}\label{defbadpacket}
An $s$-subset $\mathcal{U}\subseteq \binom{\mathcal{X}}{w}$, $2\le s\le u$, is called a \textit{bad $s$-packet} if it is spanned by at most $(s-1)w$ points.
\end{definition}

Now we are going to prove the existence of good $t$-IPPS$(w,v)$ for fixed $w$ and sufficiently large $v$. We are interested in the asymptotic behavior of the size of good $t$-IPPS$(w,v)$ for fixed $t$ and $w$. We shall show that bad $s$-packets are not typical in the case $sw\ll v$, therefore their probability is rather small and then we shall prove the existence of good set systems by using some variation of the random coding technique. Note that it is reasonable to consider large $v$ and relatively small $w$. Indeed, the distributor needs a large set of base keys to accommodate amounts of authorized users, however, each authorized user is usually assigned with a limited number of base keys which are used as the user's inputs to the decryption devices. Also, the interested reader is referred to \cite{Kaba2016} for the case that $w$ is a constant fraction of $v$, which was studied via constant weight codes.

The process of deriving a lower bound for $t$-IPPS in Theorem \ref{newlowIPPS} is to first randomly choose a family of blocks from $\binom{\mathcal{X}}{w}$, and then remove one block from each bad $s$-packet, $2\le s\le u$, which may form a possible minimal forbidden configuration.

\begin{theorem}\label{newlowIPPS}
Let $w$ and $t$ be positive integers such that $t\ge 2$. Then there exists a constant $c$, depending only on $w$ and $t$, with the following property. For any sufficiently large integer $v$, there exists a $t$-IPPS$(w,v)$ with size at least $c v^{\frac{w}{u-1}}$,
that is, $I_t(w,v)\ge c v^{\frac{w}{u-1}}$.
\end{theorem}

\begin{IEEEproof}
Let $\mathcal{X}$ be a finite set of $v$ points. Let $\binom{\mathcal{X}}{w}$ be the collection of all $w$-subsets (blocks) $B\subseteq \mathcal{X}$. Form a random subset $\mathcal{B}\subseteq \binom{\mathcal{X}}{w}$ of blocks by including each block independently with probability $p$, where $0<p<1$. We will determine the value of $p$ later.

Let $V$ denote the number of blocks in $\mathcal{B}$. Clearly, $\mathbb{E}[V]=\binom{v}{w}p$.

Let $X$ denote the number of all bad $s$-packets, $2\le s\le u$, in $\mathcal{B}$.

For any $s$-subset $\mathcal{U}\subseteq\binom{\mathcal{X}}{w}$, $2\le s\le u$, let $X(\mathcal{U})$ be the indicator random variable for the event $\mathcal{U}\subseteq \mathcal{B}$. Then
\begin{equation*}
\text{Pr}\{X(\mathcal{U})\}=p^s
\end{equation*}
as all $s$ blocks in $\mathcal{U}$ must be chosen to be in $\mathcal{B}$.

So by the linearity of expectation,
\begin{equation}\label{(1)}
\begin{split}
\mathbb{E}[X]&= \sum_{\mathcal{U}\ \text{is a bad $s$-packet in $\binom{\mathcal{X}}{w}$},\atop 2\le s\le u} \text{Pr}\{X(\mathcal{U})\}\\
&=\sum_{2\le s\le u}N_sp^{s},
\end{split}
\end{equation}
where $N_s$ is the number of bad $s$-packets in $\binom{\mathcal{X}}{w}$.

For each $2\le s\le u$, we have
\begin{equation}\label{(2)}
N_s\le \binom{v}{(s-1)w}\binom{\binom{(s-1)w}{w}}{s}.
\end{equation}
Indeed, since each bad $s$-packet is spanned by at most $(s-1)w$ points, so any $(s-1)w$ points in $\mathcal{X}$ may generate up to $\binom{\binom{(s-1)w}{w}}{s}$ bad $s$-packets in $\mathcal{X}$. There are $\binom{v}{(s-1)w}$ distinct subsets of size $(s-1)w$ in $\mathcal{X}$. The inequality (\ref{(2)}) for $N_s$ follows.

From (\ref{(1)}) and (\ref{(2)}),
\begin{equation*}
\mathbb{E}[X]\le \sum_{2\le s\le u}\binom{v}{(s-1)w}\binom{\binom{(s-1)w}{w}}{s}p^{s}.
\end{equation*}

So again by the linearity of expectation,
\begin{equation}\label{(3)}
\begin{split}
\mathbb{E}[V-X]&=\mathbb{E}[V]-\mathbb{E}[X]\\
&\ge \binom{v}{w}p-\sum_{2\le s\le u}\binom{v}{(s-1)w}\binom{\binom{(s-1)w}{w}}{s}p^{s}.
\end{split}
\end{equation}

Take $p=c_0v^{\frac{(2-u)w}{u-1}}$, where $c_0$ is a constant chosen appropriately and depending only on $w$ and $t$.
Note that for fixed $w,\ t$ and sufficiently large $v$, the value of $p$ always can be chosen such that $0<p<1$. Since $t,w,u$ are fixed, then it follows from
 (\ref{(3)}) that for  sufficiently large $v$
\begin{equation*}
\begin{split}
\mathbb{E}[V-X]&\ge c_1v^{w}p-c_2v^{(u-1)w}p^{u}-c_3\sum_{2\le s\le u-1}v^{(s-1)w}p^{s}\\
&\ge c_0c_1v^{w}v^{\frac{(2-u)w}{u-1}}-c_0^uc_2v^{(u-1)w}v^{\frac{(2-u)uw}{u-1}}-c_0^{u-1}c_3\\
&\ge cv^{\frac{w}{u-1}},
\end{split}
\end{equation*}
where $c_1,c_2,c_3$ and $c$ are constants depending only on $w$ and $t$.

Thus there exists at least one point in the probability space for which the difference $V-X$ is at least $cv^{\frac{w}{u-1}}$. That is, there is a family of blocks $\mathcal{B}$ which has at least $cv^{\frac{w}{u-1}}$ more blocks than bad $s$-packets, $2\le s\le u$. Delete one block from each bad $s$-packets, $2\le s\le u$, in $\mathcal{B}$, leaving a set $\mathcal{B}'$. This set $\mathcal{B}'$ contains no bad $s$-packet, nor minimal forbidden configuration containing $s$ blocks, $2\le s\le u$, and has at least $cv^{\frac{w}{u-1}}$ blocks.

Thus the theorem follows by Corollary \ref{sufficientIPPS}.
\end{IEEEproof}

\subsection{Remark}
\label{subsec-IPPS-remark}

In the literature, the best known upper bound for IPPS is as follows.

\begin{theorem}[\cite{GM2016}]\label{newupboundIPPS}
Let $v\ge w\ge 2,\ t\ge 2$ be integers. Then
\begin{equation*}
I_t(w,v)\le \binom{v}{\lceil\frac{w}{\lfloor t^2/4\rfloor+t}\rceil}=O(v^{\lceil\frac{w}{\lfloor t^2/4\rfloor+t}\rceil}).
\end{equation*}
\end{theorem}

In \cite{GM2016}, it was conjectured that Theorem \ref{newupboundIPPS} provides the exact upper bound for $t$-IPPS$(w,v)$, up to a constant depending only on $w$ and $t$. The lower bound for $t$-IPPS$(w,v)$ in Theorem \ref{newlowIPPS} confirms their conjecture for certain cases. That is, when $\lfloor t^2/4\rfloor+t$ is a divisor of $w$ and $v$ is large, the probabilistic construction in Theorem \ref{newlowIPPS} produces $t$-IPPS$(w,v)$ with size $cv^{\frac{w}{u-1}}$, which has the same (optimal) order of magnitude $\frac{w}{u-1}=\frac{w}{\lfloor t^2/4\rfloor+t}$ as that of the upper bound in Theorem \ref{newupboundIPPS}.

We remark that when $v$ is large, even if $\lfloor t^2/4\rfloor+t$ is not a divisor of $w$, the lower bound for $t$-IPPS$(w,v)$ in Theorem \ref{newlowIPPS} has order of magnitude $\frac{w}{\lfloor t^2/4\rfloor+t}$, which is extremely close to the order of magnitude $\lceil\frac{w}{\lfloor t^2/4\rfloor+t}\rceil$ of the upper bound  in Theorem \ref{newupboundIPPS}. Also very recently, Shangguan and Tamo \cite{ST} slightly improved this asymptotic lower bound for certain $w$ and $t$ by connecting IPPS with sparse hypergraphs.

\subsection{$2$-IPPS with small $w$}
In the preceding subsection we provided a probabilistic lower bound for the maximum size of $t$-IPPS$(w,v)$ with general $t$ and $w$. In this subsection we concentrate on the case of $t=2$, small $w$, and $v$ is any integer no less than $w$. We aim to determine the exact value of $I_t(w,v)$ in certain cases.

First we have the following lemma for a $2$-IPPS$(w,v)$, which is essentially the same as Lemma 1 in \cite{Hollmann1998}, and we omit its verification here.

\begin{lemma}\label{2IPPSabc}
A $(w,v)$ set system $(\mathcal{X},\mathcal{B})$ is a $2$-IPPS$(w,v)$ if and only if the following cases hold.
\begin{description}
  \item[(IPPSa)]\quad For any three distinct blocks $A,B,C\in \mathcal{B}$, we have
  \begin{equation*}
  |(A\cup B)\cap (A\cup C)\cap (B\cup C)|< w.
  \end{equation*}
  \item[(IPPSb)]\quad For any four distinct blocks $A,B,C,D\in \mathcal{B}$, we have
  \begin{equation*}
  |(A\cup B)\cap (C\cup D)|< w.
  \end{equation*}
\end{description}
\end{lemma}

By virtue of a construction for traceability schemes in \cite{GM2016}, we have

\begin{lemma}\label{low2IPPS}
For any $v\ge w\ge 2,\ t\ge 2$, we have
\begin{equation*}
I_t(w,v)\ge v-w+1.
\end{equation*}
\end{lemma}

\begin{IEEEproof}
We provide a construction as in \cite{GM2016}. Suppose $\mathcal{X}$ is the set of points such that $|\mathcal{X}|=v$. Arbitrarily choose a subset $\Delta\subseteq \mathcal{X}$ such that $|\Delta|=w-1$. Define
\begin{equation*}
B_j:=\{j\}\cup \Delta\subseteq \mathcal{X},\  \ \forall j\in \mathcal{X}\setminus \Delta,
\end{equation*}
and denote $\mathcal{B}:=\{B_j:\ j\in \mathcal{X}\setminus \Delta\}$. Then $(\mathcal{X},\mathcal{B})$ is a $t$-IPPS$(w,v)$ for any $t\ge 2$, since besides the common subset $\Delta$, each block possesses a unique point. The lemma follows.
\end{IEEEproof}

The following corollary follows from Lemma \ref{low2IPPS} and Theorem \ref{newupboundIPPS}.
\begin{coro}\label{IPPS23}
For any $v\ge w\ge 2,\ t\ge 2$ such that $w\le \lfloor t^2/4\rfloor+t$, we have
\begin{equation*}
v-w+1\le I_t(w,v)\le v.
\end{equation*}
\end{coro}

Now we start from the case $w=2$.

\begin{theorem}\label{IPPS2}
For any $v\ge  2$ and $t\ge 2$, we have $I_t(2,v)= v-1$.
\end{theorem}

\begin{IEEEproof}
From Corollary \ref{IPPS23}, we have $v-1\le I_t(2,v)\le v$. Since $I_t(2,v)\le I_2(2,v)$ for any $t\ge 2$, it suffices to prove that $I_2(2,v)< v$. Suppose not, and $(\mathcal{X},\mathcal{B})$ is a $2$-IPPS$(2,v)$ with size $|\mathcal{B}|=v$. We aim to find a contradiction to the definition of IPPS.

First there exists one point $x\in \mathcal{X}$ such that $x$ appears in at least two distinct blocks, since if not, then $|\mathcal{B}|\le v/2$, a contradiction to the hypothesis that $|\mathcal{B}|=v$. Denote $b(x)=|\{B\in \mathcal{B}:\ x\in B\}|$. We have $2\le b(x)\le v-1$. Notice that there exists another point $y\in \mathcal{X}\setminus \{x\}$ such that $y$ appears in at least two blocks, since $\lceil\frac{2v-b(x)}{v-1}\rceil=2$. Without loss of generality, we assume that $A,B\in \mathcal{B}$ are two distinct blocks containing $x$, and $C,D\in \mathcal{B}$ are two distinct blocks containing $y$. Note that $\{A,B\}\ne \{C,D\}$.
If $|\{A,B,C,D\}|= 3$, then, without loss of generality, assume $A=C=\{x,y\}$. It implies $A\subseteq B\cup D$, a contradiction to Lemma \ref{2IPPSabc} (IPPSa).
If $|\{A,B,C,D\}|=4$, then $\{x,y\}\subseteq (A\cup C)\cap (B\cup D)$, a contradiction to Lemma \ref{2IPPSabc} (IPPSb). Thus $I_2(2,v)< v$ and the lemma follows.
\end{IEEEproof}

For $w\ge 3$, we have an observation.
\begin{proposition}\label{sizew-1}
Let $v\ge w\ge 3$ and $(\mathcal{X},\mathcal{B})$ be a $2$-IPPS$(w,v)$. If there exist two distinct blocks $A,B\in \mathcal{B}$ such that $|A\cap B|=w-1$, then $|\mathcal{B}|\le v-w+1$.
\end{proposition}

\begin{IEEEproof}
With the assumption, we first claim that for any point $x\in \mathcal{X}\setminus (A\cap B)$, $x$ is contained in at most one block in $\mathcal{B}$. Suppose not, then there exists one point $x_0\in \mathcal{X}\setminus (A\cap B)$ contained in two blocks $C,D\in \mathcal{B}$. Clearly, $\{A,B\}\ne \{C,D\}$. If $|\{A,B,C,D\}|= 3$, then, without loss of generality, assume $A=C$. It implies that any two blocks of $A,B,D$ can generate a $w$-subset $(A\cap B)\cup\{x_0\}$, a contradiction to Lemma \ref{2IPPSabc} (IPPSa). If $|\{A,B,C,D\}|= 4$, then $(A\cap B)\cup\{x_0\}\subseteq (A\cup C)\cap (B\cup D)$, a contradiction to Lemma \ref{2IPPSabc} (IPPSb). Thus the claim follows.

Based on the above claim, $\mathcal{B}$ with the maximum number of blocks is from the construction in Lemma \ref{low2IPPS}, where $\Delta=A\cap B$. It implies $|\mathcal{B}|\le v-w+1$, as desired.
\end{IEEEproof}

Here we remark that if one would like to explore the exact value of $I_2(w,v)$, $v\ge w\ge 3$, the first step might be required to analyze the set systems with block size $w$ and
\begin{equation*}
\max\{|B_1\cap B_2|: \, B_1,B_2\in \mathcal{B},\, B_1\ne B_2\}\le w-2.
\end{equation*}

\subsection{An upper bound for 2-IPPS$(4,v)$}
\label{2-IPPS-graph}

From Section \ref{subsec-IPPS-remark}, we know that for fixed $w$, $t$ such that $(\lfloor t^2/4\rfloor+t)|w$, $I_t(w,v)=\Theta\left(v^{\frac{w}{\lfloor t^2/4\rfloor+t}}\right)$. However when $(\lfloor t^2/4\rfloor+t) \nmid w$, the order of magnitude of $I_t(w,v)$ is shown to be between $\frac{w}{\lfloor t^2/4\rfloor+t}$ and $\lceil\frac{w}{\lfloor t^2/4\rfloor+t} \rceil$. Thus it is interesting and desirable to determine the exact order of magnitude of $I_t(w,v)$ when $(\lfloor t^2/4\rfloor+t) \nmid w$. In this subsection we consider the first case $t=2$ and $w=4$. By Theorems \ref{newlowIPPS} and \ref{newupboundIPPS}, for sufficiently large $v$, we have
\begin{equation*}
c v^{4/3}\le I_2(4,v)\le \frac{1}{2}v^{2},
\end{equation*}
where $c$ is a positive constant. In the sequel, we will prove
\vskip 0.3cm
\begin{theorem}\label{main_thm}
$I_2(4,v)=o(v^2)$.
\end{theorem}

Before proving Theorem \ref{main_thm}, we do some preparations. Let $(\mathcal{X},\mathcal{B})$ be a $(w,v)$ set system and $B\in \mathcal{B}$. A subset $E\subseteq B$ is called an \textit{$|E|$-own-subset} of $B$ if for any $B'\in \mathcal{B}\setminus \{B\}$, we have $E\nsubseteq B'$. We have the following observations.

\begin{proposition}\label{1-own}
In a set system $(\mathcal{X},\mathcal{B})$, the number of blocks in $\mathcal{B}$ that contain at least one $1$-own-subset is at most $v$.
\end{proposition}

The above proposition shows that for any set system $(\mathcal{X},\mathcal{B})$, one can remove at most $v$ blocks from $\mathcal{B}$ to satisfy that each of the remaining blocks in $\mathcal{B}$ does not contain any $1$-own-subset. Next we focus on the case that a set system $(\mathcal{X},\mathcal{B})$ such that
\begin{equation}\label{assume}
\begin{split}
&\bullet\ \ \max\{|B_1\cap B_2|: \, B_1,B_2\in \mathcal{B},\, B_1\ne B_2\}\le 2,\\
&\bullet\ \ \text{for any $B\in \mathcal{B}$ and any $x\in B$, there exists $B'\in \mathcal{B}\setminus \{B\}$ such that $x\in B'$}.
\end{split}
\end{equation}

\begin{proposition}\label{distinct}
Let $(\mathcal{X},\mathcal{B})$ be a $2$-IPPS$(4,v)$. If there exist two distinct blocks $B_1,B_2\in \mathcal{B}$ such that $|B_1\cap B_2|=2$, then there does not exist $B'\in \mathcal{B}\setminus \{B_1,B_2\}$ such that $B_i\setminus (B_1\cap B_2)\subseteq B'$, where $i=1,2$.
\end{proposition}

\begin{IEEEproof}
Suppose not, without loss of generality, we assume that there exists $B'\in \mathcal{B}\setminus \{B_1,B_2\}$ such that $B_1\setminus (B_1\cap B_2)\subseteq B'$. Then $B_1\subseteq (B_2\cup B')$, a contradiction to Lemma \ref{2IPPSabc} (IPPSa).
\end{IEEEproof}

Moreover, we have
\begin{lemma}\label{no_inter_2}
Let $(\mathcal{X},\mathcal{B})$ be a $2$-IPPS$(4,v)$ such that (\ref{assume}) is satisfied.
Then
\begin{equation}\label{2_le_v-1}
|\{B\in \mathcal{B}:\ \exists\, B'\in \mathcal{B}\setminus \{B\}\ \text{such that}\ |B\cap B'|=2\}|\le v-1.
\end{equation}
\end{lemma}

\begin{IEEEproof}
Suppose on the contrary that the left-hand side of (\ref{2_le_v-1}) is no less than $v$. Then we would like to find a contradiction to that $(\mathcal{X},\mathcal{B})$ is a $2$-IPPS$(4,v)$.

To this end, we construct a graph $G=(\mathcal{X},\mathcal{E})$, where $\mathcal{X}$ is the vertex-set and $\mathcal{E}$ is the edge-set. For each $B\in \{B\in \mathcal{B}:\ \exists\, B'\in \mathcal{B}\setminus \{B\}\ \text{s.t.}\ |B\cap B'|=2\}$, the $2$-subset $B\setminus (B\cap B')$ forms an edge in $\mathcal{E}$,
where $B'$ is a block in $\mathcal{B}\setminus \{B\}$ such that $|B\cap B'|=2$. Note that a block $B$ may contribute more than one edge to $\mathcal{E}$. From Proposition \ref{distinct}, we have that any edge in $\mathcal{E}$ only belongs to one block in $\mathcal{B}$ and any two edges in $\mathcal{E}$ arising from the same block are adjacent. According to the assumption that the left-hand side of (\ref{2_le_v-1}) is not less than $v$, we have $|\mathcal{E}|\ge v$. That is, $G$ is a graph on a $v$-vertex-set containing more than $v-1$ edges. Hence there exists a cycle in $G$. Correspondingly, there exists a path of length $3$ as Fig. \ref{path3}.
\begin{figure}[htbp]
\centering
\includegraphics[scale=0.3]{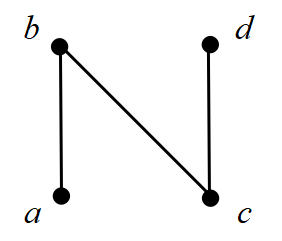}
\caption{A path of length $3$}\label{path3}
\end{figure}
If $a=d$, the path is a cycle of length $3$. The following argument is divided into two cases.

\textit{Case 1.}
If $a,b,c$ belong to the same block $B\in \mathcal{B}$, then let $B=\{a,b,c,v_1\}$, where $v_1\in \mathcal{X}\setminus \{a,b,c\}$. By the first assumption of (\ref{assume}) and the way of constructing $G$, there exist $B_1\in \mathcal{B}\setminus \{B\}$ such that $B\cap B_1=\{c,v_1\}$, and $B_2\in \mathcal{B}\setminus \{B,B_1\}$ such that $B\cap B_2=\{a,v_1\}$. Since any two blocks in $\mathcal{B}$ intersect at at most two points, there exists a point $x\in B_1$ but $x\notin B$, $x\notin B_2$. By the second assumption of (\ref{assume}), there exists a block $B_3\in \mathcal{B}\setminus \{B,B_1,B_2\}$ such that $x\in B_3$. Now we have $|\{a,c,v_1,x\}|=4$ and
\begin{equation*}
\{a,c,v_1,x\}\subseteq B\cup B_3\ \ \text{and}\ \ \{a,c,v_1,x\}\subseteq B_1\cup B_2.
\end{equation*}
However, $\{B,B_3\}$ and $\{B_1,B_2\}$ are disjoint, which implies a contradiction to Lemma \ref{2IPPSabc} (IPPSb).

Similarly, we can derive a contradiction to Lemma \ref{2IPPSabc} when $b,c,d$ belong to the same block of $\mathcal{B}$.

\textit{Case 2.}
Now we consider the case that there does not exist any block $B\in \mathcal{B}$ such that $\{a,b,c\}\subseteq B$ or $\{b,c,d\}\subseteq B$. Let $B_1=\{a,b,v_1,v_2\}\in \mathcal{B}$, where $v_1\ne v_2$, $v_1\notin \{a,b,c\}$ and $v_2\notin \{a,b,c\}$. Let $B_2,B_3$ be the blocks such that $\{b,c\}\subseteq B_2$ and $\{c,d\}\subseteq B_3$ respectively. Since $\{a,b,c\}$ is not contained in any block of $\mathcal{B}$, we have $B_1\ne B_2$ and $B_1\ne B_3$. Since $\{b,c,d\}$ is not contained in any block of $\mathcal{B}$, we have $B_2\ne B_3$. By the way of constructing $G$, there exists $B'_1\in \mathcal{B}\setminus \{B_1\}$ such that $B'_1\cap B_1=\{v_1,v_2\}$. Then $B'_1\ne B_2$ follows from the first assumption of (\ref{assume}). Now we have $|\{b,c,v_1,v_2\}|=4$ and
\begin{equation*}
\{b,c,v_1,v_2\}\subseteq B_1\cup B_2,\ \ \{b,c,v_1,v_2\}\subseteq B_1\cup B_3\ \ \text{and}\ \ \{b,c,v_1,v_2\}\subseteq B'_1\cup B_2.
\end{equation*}
If $B'_1= B_3$, then $\{b,c,v_1,v_2\}$ can be generated by any two of $B_1,B_2,B_3$, a contradiction to Lemma \ref{2IPPSabc} (IPPSa). If $B'_1\ne B_3$, then $\{B_1,B_3\}$ and $\{B'_1,B_2\}$ are disjoint but both of them can generate $\{b,c,v_1,v_2\}$, a contradiction to Lemma \ref{2IPPSabc} (IPPSb).

This completes the proof.
\end{IEEEproof}

Now we are ready to prove Theorem \ref{main_thm}, and we need the following graph removal lemma \cite{Alon1992}. In the literature, Alon \textit{et al.} \cite{Alon2001,Alon2004} used Lemma \ref{Alon} to argue the upper bounds for codes with the identifiable parent property.

\begin{lemma}[\cite{Alon1992}]\label{Alon}
For every $\gamma>0$ and every positive integer $k$, there exists a constant $\delta=\delta(k,\gamma)>0$ such that every graph $G$ on $n$ vertices, containing less than $\delta n^k$ copies of the complete graph $K_k$ on $k$ vertices, contains a set of less than $\gamma n^2$ edges whose deletion destroys all copies of $K_k$ in $G$.
\end{lemma}

\begin{IEEEproof}[Proof of Theorem \ref{main_thm}]
Proving Theorem \ref{main_thm} is equivalent to proving that for any $\epsilon>0$, there exists $v_0=v_0(\epsilon)$ such that for any $v>v_0$, we have $I_2(4,v)< \epsilon v^2$.

Suppose $(\mathcal{X},\mathcal{B})$ is a $2$-IPPS$(4,v)$ of size $M$. If there exist two distinct blocks $B_1,B_2\in \mathcal{B}$ such that $|B_1\cap B_2|=3$,
then Proposition \ref{sizew-1} ensures that $M\le v-3$, which implies $I_2(4,v)=o(v^2)$. So we only need to consider the case that $\max\{|B_1\cap B_2|: \, B_1,B_2\in \mathcal{B},\, B_1\ne B_2\}\le 2$.

First, by Proposition \ref{1-own} and Lemma \ref{no_inter_2}, we can remove at most $2v$ blocks from $\mathcal{B}$ to make the remaining $\mathcal{B}'$ satisfy that
any block does not contain any $1$-own-subset and any two blocks intersect at at most one point. Clearly, $|\mathcal{B}'|\le |\mathcal{B}|=M$ and $|\mathcal{B}'|\ge |\mathcal{B}|-2v=M-2v$.

Now we construct a graph $G=(\mathcal{X},\mathcal{E})$ by the following way: for each $B\in \mathcal{B}'$, any $2$-subset of $B$ forms an edge in $\mathcal{E}$. Obviously, any block contributes $\binom{4}{2}=6$ edges, which actually form a copy of the complete graph $K_4$. Since any two distinct blocks in $\mathcal{B}'$ intersect at at most one point, any two copies of $K_4$ in $G$ that arise from two distinct blocks are edge-disjoint. Hence, $|\mathcal{E}|\ge 6(M-2v)$. Accordingly,
one needs to delete at least $M-2v$ edges from $\mathcal{E}$ to destroy all copies of the complete graph $K_4$ in $G$.

Assume that for sufficiently large $v$, we have $M\ge \epsilon v^2$ for some $\epsilon>0$. Then we need to delete at least $M-2v\ge \frac{\epsilon}{2}v^2$ edges from $\mathcal{E}$ to destroy all copies of $K_4$ in $G$. By Lemma \ref{Alon}, let $k=4$ and $\gamma=\frac{\epsilon}{2}$, we should have that $G$ contains at least $\delta v^4$ copies of $K_4$, where $\delta=\delta(\epsilon)$ is a positive constant.

For these copies of $K_4$ in $G$, the number of copies of $K_4$ which contain at least two edges arising from the same block in $\mathcal{B}'$ is $O(v^3)$. Indeed, by an upper bound of $2$-IPPS$(4,v)$ in Theorem \ref{newupboundIPPS} that $M\le \binom{v}{2}$, there are at most $\binom{v}{2}$ ways to choose a block in $\mathcal{B}'$, and $\binom{6}{2}=15$ ways to choose two edges from that block. The above process decides at least three vertices and there are at most $v-3$ ways to choose another vertex to form a copy of $K_4$. Thus $O(v^3)$ follows.

Since $G$ contains at least $\delta v^4$ copies of $K_4$, there exists a copy of $K_4$ in which any two edges come from two different blocks. Denote one such copy of $K_4$ as Fig. \ref{K4}.
\begin{figure}[htbp]
\centering
\includegraphics[scale=0.3]{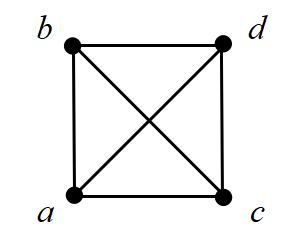}
\caption{A copy of $K_4$}\label{K4}
\end{figure}
We can suppose that $B_1,B_2,B_3,B_4\in \mathcal{B}'$ are four distinct blocks such that $\{a,b\}\subseteq B_1$, $\{a,c\}\subseteq B_2$, $\{b,d\}\subseteq B_3$ and $\{c,d\}\subseteq B_4$. Now we have $|\{a,b,c,d\}|=4$ and
\begin{equation*}
\{a,b,c,d\}\subseteq B_1\cup B_4\ \ \text{and}\ \ \{a,b,c,d\}\subseteq B_2\cup B_3.
\end{equation*}
However, $\{B_1,B_4\}$ and $\{B_2,B_3\}$ are disjoint, which implies a contradiction to Lemma \ref{2IPPSabc} (IPPSb).

Thus for any $\epsilon>0$ and sufficiently large $v$, we have $M< \epsilon v^2$, as desired.
\end{IEEEproof}

We remark that Theorem \ref{main_thm} can be generalized to the case $t=3$ and $w=6$ by a similar argument. Note that for any $t\ge 2$, $w=\lfloor (\frac{t}{2}+1)^2\rfloor$ is the smallest $w$ such that $w>t$ and $(\lfloor t^2/4\rfloor+t) \nmid w$. However, for $t\ge 4$ and $w=u=\lfloor(\frac{t}{2}+1)^2\rfloor$, we might not be able to have a similar argument as that of Theorem \ref{main_thm}. Since in a graph, we can only get $2t$ points from $t$ distinct edges, and the fact $w=\lfloor(\frac{t}{2}+1)^2\rfloor>2t$ for any $t\ge 4$ implies that $2t$ points are not enough to form a $w$-subset. But we believe that this obstacle can be removed by virtue of hypergraphs or some elaborate analyses. Precisely, we have the following conjecture.

\begin{conjecture}
Suppose $t\ge 4$ is a positive integer, then
\begin{equation*}
I_t(w,v)=o(v^2),
\end{equation*}
where $w=\lfloor(\frac{t}{2}+1)^2\rfloor$.
\end{conjecture}

\section{Probabilistic existence result for MIPPC}
\label{MIPPC}
\subsection{Multimedia parent-identifying code}

Fingerprinting in multimedia scenario was introduced in \cite{Trapper2003}, and later further developed in \cite{Blackburn2015,Cheng2011} and many other publications. In this case, the set of all authorized users' fingerprints is a code $\mathcal{C}\subseteq Q^n$. A pirate copy, generated by at most $t$ dishonest users $\mathcal{C}'\subseteq \mathcal{C}$, can reflect the information of all traitors in $\mathcal{C}'$, that is,
\begin{equation*}
f_m(\mathcal{C}')=\{\text{desc}(\mathcal{C}')\}=\{\mathcal{C}'(1)\times \mathcal{C}'(2)\times\cdots\times\mathcal{C}'(n)\},
\end{equation*}
where
\begin{equation*}
\mathcal{C}'(i)=\{\mathbf{c}(i)\in Q:\ \mathbf{c}=(\mathbf{c}(1),\ldots,\mathbf{c}(n))\in \mathcal{C}'\}.
\end{equation*}

The notion of codes with the identifiable parent property for multimedia fingerprinting was proposed in \cite{Cheng2017} for resisting the averaging attack, which was claimed as one of the most feasible collusion attacks in \cite{Wu2004}.

\begin{definition}\label{defMIPPC}
An $(n,q)$ code $\mathcal{C}$ has the \textit{$t$-identifiable parent property for multimedia fingerprinting}, denoted as $t$-MIPPC$(n,q)$, if for any subcode $\mathcal{C}'\subseteq \mathcal{C}$ such that $|\mathcal{C}'|\le t$, we have
\begin{equation*}
\bigcap_{\mathcal{S}\in S_t(\mathcal{C}')}\mathcal{S}\ne \emptyset,
\end{equation*}
where
\begin{equation*}
S_t(\mathcal{C}')=\{\mathcal{S}\subseteq \mathcal{C}:\ |\mathcal{S}|\le t,\, \text{desc}(\mathcal{S})=\text{desc}(\mathcal{C}')\}.
\end{equation*}
\end{definition}

The cardinality of $\mathcal{C}$ is called the \textit{size} of this $t$-MIPPC$(n,q)$. Denote $M_t(n,q)$ as the maximum size of a $t$-MIPPC$(n,q)$. A $t$-MIPPC$(n,q)$ $\mathcal{C}$ is \textit{optimal} if it has size $M_t(n,q)$. The \textit{code rate} of a $t$-MIPPC$(n,q)$ with size $M$ is
\begin{equation*}
R_q(n,t)=\frac{\log_q{M}}{n},
\end{equation*}
and it is \textit{optimal} provided $M=M_t(n,q)$.

The above MIPPCs is a variation of IPP codes introduced by Hollmann, van Lint, Linnartz and Tolhuizen \cite{Hollmann1998}. Also from Definition \ref{defIPPscheme} and formula (\ref{eq-f-IPP}), we state the definition of IPP codes as follows.

\begin{definition}\label{def-IPPC}
An $(n,q)$ code $\mathcal{C}$ has the \textit{$t$-identifiable parent property}, denoted as $t$-IPPC$(n,q)$, if for any subcode $\mathcal{C}'\subseteq \mathcal{C}$ such that $|\mathcal{C}'|\le t$, and any $d\in \text{desc}(\mathcal{C}')$, we have
\begin{equation*}
\bigcap_{\mathcal{P}\in P_t(d)}\mathcal{P}\ne \emptyset,
\end{equation*}
where
\begin{equation*}
P_t(d)=\{\mathcal{P}\subseteq \mathcal{C}:\ |\mathcal{P}|\le t,\, d\in \text{desc}(\mathcal{P}) \}.
\end{equation*}
\end{definition}

There is a relationship between MIPPC and IPPC.
\begin{proposition}[\cite{Cheng2017}]
A $t$-IPPC$(n,q)$ is a $t$-MIPPC$(n,q)$.
\end{proposition}
Based on this relationship and the known results on IPPC in \cite{Barg2001,Blackburn2003IPP}, we could have a lower bound for MIPPC. Namely,
\begin{equation}\label{eq-lower-IPPC}
\lim_{q\to \infty} R_q(n,t)\ge \frac{1}{\lfloor t^2/4\rfloor+t}.
\end{equation}
In the next subsection, we will show a much better lower bound than (\ref{eq-lower-IPPC}) for MIPPC.

\subsection{A lower bound for $t$-MIPPC}

From Definition \ref{defforbiddenconfiguration}, we have the following description of forbidden configurations in a $t$-MIPPC. Let $\mathcal{C}$ be a $t$-MIPPC,
a (minimal) \textit{forbidden} configuration in $\mathcal{C}$ is a (minimal) configuration $\mathcal{F}=\{\mathcal{F}_1,\ldots,\mathcal{F}_m\}$, $\mathcal{F}_i\subseteq \mathcal{C},\ |\mathcal{F}_i|\le t$, $i=1,\ldots,m$, such that
\begin{equation*}
\text{desc}(\mathcal{F}_1)=\text{desc}(\mathcal{F}_2)=\cdots =\text{desc}(\mathcal{F}_m),
\end{equation*}
that is, for each $1\le i\le n$,
\begin{equation*}
\mathcal{F}_1(i)=\mathcal{F}_2(i)=\cdots=\mathcal{F}_m(i).
\end{equation*}

By Proposition \ref{relationshipIPPscheme}, we have the following corollary.
\begin{coro}\label{sufficientMIPPC}
An $(n,q)$ code $\mathcal{C}$ is a $t$-MIPPC$(n,q)$ if and only if there are no minimal forbidden configurations in $\mathcal{C}$ with size at most $u$.
\end{coro}

To use the expurgation method to derive a lower bound for $t$-MIPPC$(n,q)$ in Theorem \ref{lowerMIPPC}, first, we randomly choose some words from $Q^n$, next, we try to compute the expected number of the possible minimal forbidden configurations with size at most $u$ and delete one word from each of these possible minimal forbidden configurations.

\begin{theorem}\label{lowerMIPPC}
Let $n$ and $t$ be positive integers such that $n\ge 2, t\ge 2$. Then there exists a constant $c$, depending only on $n$ and $t$, with the following property. For any sufficiently large integer $q$, there exists a $t$-MIPPC$(n,q)$ with size $cq^{\frac{tn}{2t-1}}$, that is, $M_t(n,q)\ge cq^{\frac{tn}{2t-1}}$.
\end{theorem}

\begin{IEEEproof}
Let $Q=\{0,1,\ldots,q-1\}$ be a set of cardinality $q$. Choose words $\mathbf{c}_1,\mathbf{c}_2,\ldots,$ $\mathbf{c}_M\in Q^n$ uniformly and independently at random, where $M$ is an integer to be decided later. Denote $\mathcal{C}:=\{\mathbf{c}_1,\mathbf{c}_2,\ldots,\mathbf{c}_M\}$.

Now we would like to remove some words from $\mathcal{C}$ to avoid the forbidden configurations of size at most $u$. To this end, we first define the \textit{bad subfamily} as follows. A subfamily $\mathcal{U}\subseteq \mathcal{C}$ is called \textit{bad} if there exist $m$ subsets of $\mathcal{U}$, say $\mathcal{F}_1,\ldots,\mathcal{F}_m$, such that
\begin{itemize}\label{def-badcode}
  \item[(a)] $\mathcal{U}=\bigcup_{1\le i\le m}\mathcal{F}_i$ and $|\mathcal{F}_i|\le t$ for any $1\le i\le m$;
  \item[(b)] $\bigcap_{1\le i\le m}\mathcal{F}_i=\emptyset$;
  \item[(c)] for any $1\le i\le n$, $\mathcal{F}_1(i)=\mathcal{F}_2(i)=\cdots=\mathcal{F}_m(i)$.
\end{itemize}
Note that the bad subfamilies here are different from the (minimal) forbidden configurations. From the natural setting, forbidden configurations cannot be multi-sets while a bad subfamily might be a multi-subset of $Q^n$.

Now we are going to compute the expected number of bad subfamilies in $\mathcal{C}$ with size at most $u$ and remove one word from each bad subfamily to obtain a $t$-MIPPC. It is clear that $u\ge 2t$. The following process will be divided into two cases according to the size of a bad subfamily: (1) the size is less than $2t$; (2) the size is no less than $2t$.

\textit{Case 1.}
We first consider the case when the size of a bad subfamily is less than $2t$. Suppose $\mathcal{U}=\{\mathbf{f}_1,\ldots,\mathbf{f}_{\delta}\}\subseteq \mathcal{C}$
is a bad subfamily, where $2\le \delta< 2t$. Then for each coordinate $1\le i\le n$, we have a useful observation, that is,
\begin{equation}\label{(4)}
|\mathcal{U}(i)|=|\{\mathbf{f}_1(i),\ldots,\mathbf{f}_{\delta}(i)\}|\le \frac{\delta}{2}. 
\end{equation}
Indeed, for each $\mathbf{f}_j(i)$, $1\le j\le \delta$, there exists another word $\mathbf{f}_k\in \mathcal{U}\setminus \{\mathbf{f}_j\}$ such that $\mathbf{f}_k(i)=\mathbf{f}_j(i)$. If not, without loss of generality, we may assume that $\mathbf{f}_1(i)\ne \mathbf{f}_k(i)$ for any $2\le k\le \delta$. Since $\mathcal{U}$ is a bad subfamily, there exist $m$ subsets $\mathcal{F}_1,\ldots,\mathcal{F}_m$ of $\mathcal{U}$ satisfying conditions (a), (b) and (c). By condition (a), there exists a subset $\mathcal{F}_{s}$, $1\le s\le m$, such that $\mathbf{f}_1\in \mathcal{F}_{s}$. Accordingly, $\mathbf{f}_1(i)\in \mathcal{F}_{s}(i)$. By
condition (c), for any $\mathcal{F}_{j}$, $1\le j\le m$, we have $\mathcal{F}_{j}(i)=\mathcal{F}_{s}(i)$. Thus $\mathbf{f}_1(i)\in \mathcal{F}_{j}(i)$, which implies $\mathbf{f}_1\in \mathcal{F}_{j}$. Accordingly,
\begin{equation*}
\mathbf{f}_1\in \bigcap_{1\le j\le m}\mathcal{F}_{j}\ne \emptyset,
\end{equation*}
which does not satisfy condition (b) and thus contradicts that $\mathcal{U}$ is a bad subfamily. Hence (\ref{(4)}) follows. Actually, the right-hand side of inequality (\ref{(4)}) should be $\lfloor \frac{\delta}{2}\rfloor$, but we omit the floor-function symbol for convenience.

Based on the above observation, we estimate the probability of the event that a given $\delta$-subfamily of $\mathcal{C}$ forms a bad subfamily. In the following estimation, we always consider the case that $q$ is much larger than $n$ and $t$. Let $\mathcal{U}\subseteq \mathcal{C}$ such that $|\mathcal{U}|=\delta$, $2\le \delta< 2t$. Clearly, there are $\binom{M}{\delta}$ distinct such $\mathcal{U}$ in $\mathcal{C}$. Let $X(\mathcal{U})$ be the event that $\mathcal{U}$ forms a bad subfamily. For each $1\le i\le n$, the $i$-th coordinates $\mathcal{U}(i)$ contributes to the event $X(\mathcal{U})$ with the probability at most
\begin{equation*}
\binom{q}{{\delta}/{2}}(\delta/2)^{\delta}/{q^{\delta}}.
\end{equation*}
Since each coordinate contributes to the event $X(\mathcal{U})$ independently, we have
\begin{equation*}
\text{Pr}\{X(\mathcal{U})\}\le \binom{q}{{\delta}/{2}}^n(\delta/2)^{\delta n}/{q^{\delta n}}.
\end{equation*}
Therefore, the expectation number of bad subfamilies in $\mathcal{C}$ with size less than $2t$ is at most
\begin{equation}\label{(5)}
\sum_{2\le \delta< 2t}\binom{M}{\delta}\binom{q}{{\delta}/{2}}^n(\delta/2)^{\delta n}/{q^{\delta n}}.
\end{equation}

\textit{Case 2.} Next we consider the case when the size of a bad subfamily is no less than $2t$. Suppose $\mathcal{U}=\{\mathbf{f}_1,\ldots,\mathbf{f}_{\gamma}\}\subseteq \mathcal{C}$ is a bad subfamily, where $2t\le \gamma\le u$. Similarly, for each $1\le i\le n$, we have an observation that
\begin{equation}\label{(6)}
|\mathcal{U}(i)|=|\{\mathbf{f}_1(i),\ldots,\mathbf{f}_{\gamma}(i)\}|\le t.
\end{equation}
If not, then $|\mathcal{U}(i)|\ge t+1$. Without loss of generality, we may assume that $|\{\mathbf{f}_1(i),\ldots,\mathbf{f}_{t+1}(i)\}|=t+1$. Since $\mathcal{U}$ is a bad subfamily, there exist $m$ subsets $\mathcal{F}_1,\ldots,\mathcal{F}_m$ of $\mathcal{U}$ satisfying conditions (a), (b) and (c). By condition (a), we know that $\mathcal{F}_1(i)$ contains at most $t$ elements in $\mathcal{U}(i)$, and there exists another $\mathcal{F}_j$, $j\ne 1$, such that
\begin{equation*}
\mathcal{F}_j(i)\cap (\mathcal{U}(i)\setminus \mathcal{F}_1(i))\ne \emptyset.
\end{equation*}
However, this implies $\mathcal{F}_j(i)\ne \mathcal{F}_1(i)$, which contradicts condition (c) and implies that $\mathcal{U}$ is not a bad subfamily. Thus (\ref{(6)}) follows.

Now we are going to estimate the probability of the event that a given $\gamma$-subfamily of $\mathcal{C}$ forms a bad subfamily. Let $\mathcal{U}\subseteq \mathcal{C}$ such that $|\mathcal{U}|=\gamma$, $2t\le \gamma\le u$. Clearly, there are $\binom{M}{\gamma}$ distinct such $\mathcal{U}$ in $\mathcal{C}$. Let $X(\mathcal{U})$ be the event that $\mathcal{U}$ forms a bad subfamily. For each $1\le i\le n$, the $i$-th coordinates $\mathcal{U}(i)$ contributes to the event $X(\mathcal{U})$ with the probability at most
\begin{equation*}
\binom{q}{t}t^{\gamma}/{q^{\gamma}}.
\end{equation*}
Since each coordinate contributes to the event $X(\mathcal{U})$ independently, we have
\begin{equation*}
\text{Pr}\{X(\mathcal{U})\}\le \binom{q}{t}^nt^{\gamma n}/{q^{\gamma n}}.
\end{equation*}
Therefore, the expectation number of bad subfamilies in $\mathcal{C}$ with size $\ge 2t$ and $\le u$ is at most
\begin{equation}\label{(7)}
\sum_{2t\le \gamma\le u}\binom{M}{\gamma}\binom{q}{t}^nt^{\gamma n}/{q^{\gamma n}}.
\end{equation}

Now we form a set $\mathcal{B}$ by choosing one word from each bad subfamily in $\mathcal{C}$ of size $\ge 2$ and $\le u$. Then from (\ref{(5)}) and (\ref{(7)}), we have
\begin{equation*}
\begin{split}
|\mathcal{B}|&\le \sum_{2\le \delta< 2t}\binom{M}{\delta}\binom{q}{{\delta}/{2}}^n(\delta/2)^{\delta n}/{q^{\delta n}}+ \sum_{2t\le \gamma\le u}\binom{M}{\gamma}\binom{q}{t}^nt^{\gamma n}/{q^{\gamma n}}.
\end{split}
\end{equation*}

Define $\hat{\mathcal{C}}=\mathcal{C}\setminus \mathcal{B}$. Clearly, any two words in $\hat{\mathcal{C}}$ are distinct, since we have removed one word from each bad subfamily of size $2$. Moreover,
\begin{equation}\label{(8)}
\begin{split}
|\hat{\mathcal{C}}|&=|\mathcal{C}|-|\mathcal{B}|\\
&\ge M-\sum_{2\le \delta< 2t}\binom{M}{\delta}\binom{q}{{\delta}/{2}}^n \left(\frac{\delta}{2q}\right)^{\delta n}- \sum_{2t\le \gamma\le u}\binom{M}{\gamma}\binom{q}{t}^n
\left(\frac{t}{q}\right)^{\gamma n}.
\end{split}
\end{equation}
We claim that $\hat{\mathcal{C}}$ is a $t$-MIPPC. Since if not, by Corollary \ref{sufficientMIPPC}, there would exist a minimal forbidden configuration with size at most $u$. Correspondingly, there would exist a bad subfamily of size at most $u$. But we have already destroyed all bad subfamilies in $\mathcal{C}$ by removing one word from each of them. The definitions of (minimal) forbidden configuration and bad subfamily are inner definitions, i.e., these properties do not depend on a code in whole. Thus there does not exist any forbidden configuration with size at most $u$ in $\hat{\mathcal{C}}$. Hence $\hat{\mathcal{C}}$ is a $t$-MIPPC.

For sufficiently large $q$, let $M=\epsilon q^{\frac{tn}{2t-1}}$, where $\epsilon$ is a constant chosen appropriately and depending only on $n$ and $t$. Substituting it into (\ref{(8)}), we have
\begin{equation*}
\begin{split}
|\hat{\mathcal{C}}|
&\ge M-\sum_{2\le \delta< 2t}\binom{M}{\delta}\binom{q}{{\delta}/{2}}^n \left(\frac{\delta}{2q}\right)^{\delta n}- \sum_{2t\le \gamma\le u}\binom{M}{\gamma}\binom{q}{t}^n \left(\frac{t}{q}\right)^{\gamma n}\\
&\ge M-\kappa_1 \sum_{2\le \delta< 2t} M^{\delta}q^{-\delta n/2}-\kappa_2 \sum_{2t\le \gamma\le u} M^{\gamma}q^{(t-\gamma)n} \\
&=M-\kappa_2 M^{2t}q^{-t n}-  \kappa_1 \sum_{2\le \delta\le 2t-1} M^{\delta}q^{-\delta n/2}-\kappa_2 \sum_{2t\le \gamma\le u,\atop \gamma\ne 2t} M^{\gamma}q^{(t-\gamma)n}\\
&\ge \epsilon q^{\frac{tn}{2t-1}}-\kappa_2 \epsilon^{2t} q^{\frac{tn}{2t-1}}- \kappa_1' q^{\frac{n}{2}}-\kappa_2' q^{\frac{n}{2t-1}}\\
&\ge c q^{\frac{tn}{2t-1}},
\end{split}
\end{equation*}
where $\kappa_1,\kappa_2,\kappa_1',\kappa_2'$ and $c$ are constants depending only on $n$ and $t$. The theorem follows.
\end{IEEEproof}

\subsection{Remarks}

In \cite{Cheng2017}, Cheng \textit{et al.} transferred the requirement of a $t$-MIPPC to a corresponding bipartite graph without cycles of length $\le 2t$, obtaining the following upper bound.

\begin{theorem}[\cite{Cheng2017}]\label{upperMIPP}
Let $n,t,q$ be positive integers. Then
\begin{equation*}
M_t(n,q)\le \begin{cases}
 q^{\frac{n}{2}}(q^{\frac{n}{2t}} +2c) &\mbox{if $n$ is even}\\
 q^{\frac{n}{2}}(q^{\frac{n+1}{2t}} +c(q^{\frac{1}{2}}+q^{-\frac{1}{2}})) &\mbox{if $n$ is odd and $t$ is even}\\
 q^{\frac{n}{2}}(q^{\frac{n}{2t}} +c(q^{\frac{1}{2}}+q^{-\frac{1}{2}})) &\mbox{if $n$ is odd and $t$ is odd},
 \end{cases}
\end{equation*}
where $c$ is a constant depending only on $t$.
\end{theorem}

From Theorem \ref{lowerMIPPC} and Theorem \ref{upperMIPP}, we have

\begin{coro}\label{compare}
Let $n$ and $t$ be positive integers, then
\begin{equation*}
\frac{t}{2t-1}\le \lim_{q\to \infty}R_q(n,t)
\le \begin{cases}
 \frac{1}{2}+\frac{1}{2t} &\mbox{if $n$ is even}\\
 \frac{1}{2}+\max\{\frac{n+1}{2tn}, \frac{1}{2n}\} &\mbox{if $n$ is odd and $t$ is even}\\
 \frac{1}{2}+\max\{\frac{1}{2t}, \frac{1}{2n}\} &\mbox{if $n$ is odd and $t$ is odd}.
 \end{cases}
\end{equation*}
\end{coro}

In Table \ref{table1}, we list the comparison in Corollary \ref{compare} for certain parameters, which shows that the gap is very small.

\begin{table}\centering
\begin{tabular}{|c|c|c|c|c|c|c|c|c|c|c|c|}
  \hline
  $(n,t)$ & $(3,2)$ & $(2,3)$ & $(4,4)$ & $(6,5)$ & $(8,6)$ & $(9,7)$ & $(13,10)$ & $(15,11)$ & $(17,12)$ & $(19,13)$\\ \hline
  $\lim_{q\to \infty}R_q(n,t)\ge$ & $2/3$ & $0.6$ & $0.571$ & $0.556$ & $0.545$ & $0.538$ & $0.526$ & $0.524$ & $0.522$ & $0.52$\\ \hline
  $\lim_{q\to \infty}R_q(n,t)\le$ & $2/3$ & $0.667$ & $0.625$ & $0.583$ & $0.563$ & $0.571$ & $0.554$ & $0.545$ & $0.544$ & $0.538$\\
  \hline
\end{tabular}
\caption{Bounds for $t$-MIPPC of length $n$}\label{table1}
\end{table}

Furthermore, we compare the code rate of a $t$-MIPPC with another kind of multimedia fingerprinting code, namely, $\bar{t}$-separable code. Separable code was introduced in \cite{Cheng2011} and was studied by several authors, see \cite{Blackburn2015,Cheng2012,Egorova,GG2014,YYZ} for example.

When $t=2$, it was proved that a $2$-MIPPC is exactly a $\bar{2}$-separable code \cite{Cheng2017}. In Theorem \ref{lowerMIPPC}, for fixed $n$ such that $n\equiv 0\,(\bmod\ 3)$ and sufficiently large $q$, the asymptotic code rate of a $2$-MIPPC could be at least $2/3$, which matches the asymptotically optimal code rate of $\bar{2}$-separable code in \cite{Blackburn2015} and \cite{GG2014}. This implies that, in Theorem \ref{lowerMIPPC}, the expurgation method provides the $2$-MIPPC with asymptotically optimal code rate when $n$ is a multiple of $3$  and $q$ is large.

For fixed $n\ge 2, t\ge 3$ and sufficiently large $q$, Blackburn \cite{Blackburn2015} proved that, when $(t-1)|n$, the asymptotically optimal code rate of a $\bar{t}$-separable code is $1/(t-1)$. However, by Corollary \ref{compare}, the fact
\begin{equation*}
\frac{t}{2t-1}>\frac{1}{t-1}
\end{equation*}
implies that a $t$-MIPPC may provide much more codewords than a $\bar{t}$-separable code.

Finally, we make a comparison between MIPPC and IPPC. It was proved in \cite{Blackburn2003IPP} that for fixed $n\ge 2, t\ge 2$ such that $(\lfloor t^2/4\rfloor+t)|n$, the asymptotically optimal code rate of a $t$-IPPC$(n,q)$ is $1/(\lfloor t^2/4\rfloor+t)$. Thus Corollary \ref{compare} and the fact
\begin{equation*}
\frac{t}{2t-1}>\frac{1}{\lfloor t^2/4\rfloor+t}
\end{equation*}
show that a $t$-MIPPC has a much larger asymptotic code rate than a $t$-IPPC.

\section{Conclusion}
\label{conclusion}

In this paper we considered combinatorial structures for parent-identifying schemes. First we established an equivalent relationship between parent-identifying schemes and forbidden configurations. Based on this relationship, we derived probabilistic existence results for IPPS and MIPPC respectively. The probabilistic lower bound for the maximum size of $t$-IPPS has the asymptotically optimal order of magnitude in many cases, and that for $t$-MIPPC provides the asymptotically optimal code rate when $t=2$ and the best known asymptotic code rate when $t\ge 3$. We analyzed the structure of $2$-IPPS and proved some bounds for certain cases. However, there is still a gap between the known upper bounds and lower bounds for IPPS and MIPPC in many other cases. It would be of interest to narrow the gap and further to determine the exact maximum size (or code rate) for them.

\section*{Acknowledgments}

Y. Gu would like to thank Prof. Yuichiro Fujiwara and Dr. Chong Shangguan for stimulating discussions.

\end{document}